\documentclass[12pt]{article}

\usepackage{epsfig}

\def\be{\begin{equation}}
\def\ee{\end{equation}}
\def\ba{\begin{eqnarray}}
\def\ea{\end{eqnarray}}
\def\({\left (}
\def\){\right )}
\def\[{\left [}
\def\[{\right ]}

\begin{document}

\begin{titlepage}
\bigskip
\rightline{CERN-PH-TH/2006-022}
\rightline{hep-th/0602091}
\bigskip\bigskip
\bigskip\bigskip

\centerline{\Large \bf {Populating the Landscape: A Top Down Approach}}
\bigskip\bigskip\bigskip\bigskip
\centerline{\large S.W. Hawking\footnote{S.W.Hawking@damtp.cam.ac.uk} and 
Thomas Hertog\footnote{Thomas.Hertog@cern.ch}}
\bigskip\bigskip
\centerline{\em $^1$ DAMTP, University of Cambridge, 
                Wilberforce Road, Cambridge CB3 0WA, UK}
\bigskip
\centerline{\em $^2$ Physics Department, Theory Division, CERN, 
                CH-1211 Geneva 23, Switzerland}
\bigskip\bigskip

\begin{abstract}

We put forward a framework for cosmology that combines the string landscape with no boundary initial 
conditions. In this framework, amplitudes for alternative histories for the universe are calculated with 
final boundary conditions only. This leads to a top down approach to cosmology, in which the histories of 
the universe depend on the precise question asked. We study the observational consequences of no boundary 
initial conditions on the landscape, and outline a scheme to test the theory. This is illustrated  in a 
simple model landscape that admits several alternative inflationary histories for the universe. Only a few 
of the possible vacua in the landscape will be populated. We also discuss in what respect the top down approach 
differs from other approaches to cosmology in the string landscape, like eternal inflation.

\end{abstract}

\end{titlepage}

\baselineskip=18pt

%%%%%%%%%%%%%%%%%%%%%%%%%%%%%%
\section{Introduction}
%%%%%%%%%%%%%%%%%%%%%%%%%%%%%%

It seems likely that string theory contains a vast ensemble of stable and metastable vacua, including some with 
a small positive effective cosmological constant \cite{Bousso00} and the low energy effective field theory of the 
Standard Model. Recent progress on the construction of metastable de Sitter vacua \cite{Kachru03} lends further 
support to the notion of a string landscape \cite{Susskind03}, and a statistical analysis
gives an idea of the distribution of some properties among 
the vacua \cite{Douglas03}. But it has remained unclear what is the correct framework for cosmology in the 
string landscape. There are good reasons to believe, however, that a proper understanding of the cosmological 
dynamics will be essential for the landscape to be predictive \cite{Banks04}.

In particle physics, one usually computes S-matrix elements. This is useful to predict the outcome of 
laboratory experiments, where one prepares the initial state and measures the final state. It could be viewed 
as a bottom-up approach to physics, in which one evolves forward in time a particular initial state of the 
system. The predictivity of this approach arises from and relies upon the fact that one has control over the 
initial state, and that experiments can be repeated many times to gain statistically significant results.

But cosmology poses questions of a very different character. In our past there is an epoch of the early universe 
when quantum gravity was important. The remnants of this early phase 
are all around us. The central problem in cosmology is to understand why these remnants are what they are, and 
how the distinctive features of our universe emerged from the big bang. Clearly it is not an S-matrix that is 
the relevant observable\footnote{See \cite{Banks01,Witten01,Bousso05,Giddings05} for recent work on the existence 
and the construction of observables in cosmological spacetimes.} for these predictions, since we live in 
the middle of this particular experiment. Furthermore, we have no control over the initial state of the universe, 
and there is certainly no opportunity for observing multiple copies of the universe. 

In fact if one does adopt a bottom-up approach to cosmology, one is immediately led to an essentially classical 
framework, in which one loses all ability to explain cosmology's central question - why our universe is the way 
it is. In particular a bottom-up approach to cosmology either requires one to postulate an initial state of the 
universe that is carefully fine-tuned \cite{Veneziano91} - as if prescribed by an outside agency - or it requires 
one to invoke the notion of eternal inflation \cite{Vilenkin83}, which prevents one from predicting what a typical 
observer would see.

Here we put forward a different approach to cosmology in the string landscape, based not on the classical idea of 
a single history for the universe but on the quantum sum over histories \cite{Hartle95}. 
We argue that the quantum origin of the universe naturally leads to a framework for cosmology where amplitudes for 
alternative histories of the universe are computed with boundary conditions at late times only. 
We thus envision a set of alternative universes in the landscape, with amplitudes given by
the no boundary path integral \cite{Hartle83}. 

The measure on the landscape provided by no boundary initial conditions allows one to derive predictions for 
observations. This is done by evaluating probabilities for alternative histories that obey a set of constraints 
at late times. The constraints provide information that is supplementary to the fundamental laws and act as a
selection principle. In particular, they select the subclass of histories that contribute to the amplitude of interest.
One then identifies alternatives within this subclass that have probabilities near one. These include, in particular,
predictions of future observations. The framework we propose is thus more like a top down approach 
to cosmology, where the histories of the universe depend on the precise question asked. 

We illustrate our framework in a model landscape that admits several distinct classes of
inflationary histories for the universe. In this model, we predict several properties of the subclass of histories
that are three-dimensional, expanding and approximately flat at late times. We also discuss in general terms the 
predictions of top down cosmology in more complicated models like the string landscape. 

Finally we discuss in what respect the top down approach differs from other (bottom-up) approaches to cosmology 
in the string landscape, such as eternal inflation or pre-big bang cosmology.

%%%%%%%%%%%%%%%%%%%%%%%%%%%%%%
\section{Quantum State}
%%%%%%%%%%%%%%%%%%%%%%%%%%%%%%

In cosmology one is generally not concerned with observables at infinity or with properties of the 
entire four-geometry, but with alternatives in some finite region in the interior of the spacetime. 
The amplitudes for these more restricted sets of observables are obtained from the amplitudes of four 
dimensional metric and matter field configurations, by integrating over the unobserved 
quantities\footnote{The precise relation between familiar quasilocal observables and the 
diffeomorphism-invariant observables of quantum gravity remains an important outstanding 
issue. See e.g.\cite{Giddings05} for recent work on this.}.
A particularly important case is the amplitude of finding a compact spacelike surface $S$ 
with induced three-metric $g^3_{ij}$ and matter field configuration $\phi$,
\be\label{wavefn}
\Psi [g^3,\phi ] \sim
\int_{C} [{\cal D} g ][{\cal D} \phi] \ e^{iS[g,\phi]}.
\ee
Here the path integral is taken over the class $C$ of spacetimes which agree with $g_{ij}^3$ and
$\phi$ on a compact boundary $S$. The quantum state of the universe is determined by the
remaining specification of the class $C$.

Usually one sums over histories that have an initial and a final boundary. This is useful for the computation 
of S-matrix elements to predict the outcome of laboratory experiments, where one prepares the initial state and 
measures the final state. It is far from clear, however, that this 
is the appropriate setup for cosmology, where one has no control over the initial state, and no opportunity for 
observing multiple copies of the universe. 
In fact, if one does apply this approach to cosmology one is naturally led to an essentially classical 
picture, in which one simply assumes the universe began and evolved in a way that is well defined and 
unique. 

Pre-big bang cosmologies \cite{Veneziano91} are examples of models that are based on a bottom-up
approach. In these models one specifies an initial state on a surface in the infinite past and evolves 
this forward in time. A natural choice for the initial state would be flat space, but that would obviously 
remain flat space. Thus one instead starts with an unstable state in the infinite past, tuned carefully in 
order for the big crunch/big bang transition to be smooth and the path integral to be peaked around a single 
semi-classical history. Several explicit solutions of such bouncing cosmologies have been found in various 
minisuperspace approximations \cite{Karczmarek04}. It has been shown, however, using several
different techniques, that solutions of this kind are unstable \cite{Seiberg02,Hertog04}. 
In particular, one finds 
that generic small perturbations at early times (or merely taking in account the remaining degrees of freedom) 
dramatically change the evolution near the transition. Rather than evolving towards an expanding 
semi-classical universe at late times, one generically produces a strong curvature 
singularity. Hence the evolution of pre-big bang cosmologies always includes a genuinely
quantum gravitational phase, unless the initial state is extremely fine-tuned. It is therefore more appropriate
to describe these cosmologies by a path integral in quantum cosmology, and not in terms of a single 
semi-classical trajectory. The universe won't have a single history but every possible history, each 
with its own probability.

In fact, the quantum state of the universe at late times is likely to be independent of the state on 
the initial surface. This is because there are geometries in which the initial surface is in one universe
and the final surface in a separate disconnected universe. Such metrics exist in the Euclidean regime, 
and correspond to the quantum annihilation of one universe and the quantum creation of another. Moreover, 
because there are so many different possible universes, these geometries dominate the path integral. 
Therefore even if the path integral had an initial boundary in the infinite past, the state on a 
surface $S$ at late times would be independent of the state on the initial surface. It would be given by 
a path integral over all metric and matter field configurations whose only boundary is the final surface 
$S$. But this is precisely the no boundary quantum state \cite{Hartle83} 
\be\label{nbp}
\Psi [g^3,\phi ] \sim
\int_{C} [{\cal D} g ][{\cal D} \phi] \ e^{-S_{E}[g,\phi]},
\ee
where the integral is taken over all regular geometries bounded only by the compact 
three-geometry $S$ with induced metric $g^{3}_{ij}$ and matter field configuration $\phi$.
The Euclidean action $S_{E}$ is given by\footnote{We have set $8\pi G=1$.}
\be\label{Eact}
S_{E} =  -\frac{1}{2} \int d^4 x \sqrt{g} \left( R + L(g,\phi) \right) -\int_{S} d^3 x \sqrt{g^3} K,
\ee
where $L(g,\phi)$ is the matter Lagrangian.

One expects that the dominant contributions to the path integral will come from saddle points in the action. 
These correspond to solutions of the Einstein equations with the prescribed final boundary condition. If 
their curvature is bounded away from the Planck value, the saddle point metric will be in the semi-classical 
regime and can be regarded as the most probable history of the universe. 
Saddle point geometries of particular interest include geometries where a Lorentzian metric is rounded off 
smoothly in the past on a compact Euclidean {\em instanton}. Well known examples of such geometries are the 
Hawking--Moss (HM) instanton \cite{Hawking82a} which matches to Lorentzian de Sitter space, and the Coleman--De 
Luccia (CdL)
instanton \cite{Coleman80}, which continues to an open FLRW universe. The former occurs generically in models 
of gravity coupled to scalar fields, while the latter requires a rather fine-tuned potential.

The usual interpretation of these geometries is that they describe the decay of a false vacuum in
de Sitter space. However, they have a different interpretation in the no boundary proposal 
\cite{Gratton99}. Here they 
describe the beginning of a new, independent universe with a completely self-contained `no boundary'
description\footnote{The interpretation of these saddle point geometries is in line with their 
interpretation that
follows from holographic reasoning, as described e.g. in \cite{Dyson02}. Some of our conclusions, however, 
differ from \cite{Dyson02}.}. By this we mean, in particular, that the expectation values of observables that are 
relevant to local observers within the universe can be unambiguously computed from the no boundary path 
integral, without the need for assumptions regarding the pre-bubble era. The original de Sitter universe 
may continue to exist, but it is irrelevant for observers inside the new universe. The no boundary proposal 
indicates, therefore, that the pre-bubble inflating universe is a redundant theoretical construction.

It is appealing that the no boundary quantum state (\ref{nbp}) is computed directly from the action 
governing the dynamical laws. There is thus essentially a single theory of dynamics and of the quantum state. 
It should be emphasized however that this remains a {\it proposal} for the wave function of the universe. 
We have argued it is a natural choice, but the ultimate test is whether its predictions agree with 
observations.

%%%%%%%%%%%%%%%%%%%%%%%%%%%%%%
\section{Prediction in Quantum Cosmology}
%%%%%%%%%%%%%%%%%%%%%%%%%%%%%%

Quantum cosmology aims to identify which features of the observed universe follow directly from 
the fundamental laws, and which features can be understood as consequences of quantum accidents or 
late time selection effects. In no boundary cosmology, where one specifies boundary conditions at late 
times only, this program is carried out by evaluating probabilities for alternative histories that obey
certain constraints at the present time. The final boundary conditions provide information that is supplementary 
to the fundamental laws, which selects 
a subclass of histories and enables one to identify alternatives that (within this subclass) have probabilities 
near one. In general the probability for an alternative $\alpha$, given $H$, $\Psi$ and a set of constraints 
$\beta$, is given by
\be\label{condprob}
p(\alpha \vert \beta, H, \Psi) = \frac{ p(\alpha, \beta \vert H,\Psi)}{p(\beta \vert H, \Psi)}.
\ee
The conditions $\beta$ in (\ref{condprob}) generally contain environmental selection effects, 
but they also include features that follow from quantum accidents in the early universe\footnote{These are
quantum accidents that became `frozen', leaving an imprint on the universe at late times.}.

A typical example of a condition $\beta$ is the dimension $D$ of space. For good reasons, one usually 
considers string compactifications down to three space dimensions. However, there appears to be no dynamical
reason for the universe to have precisely four large dimensions. Instead, the no boundary proposal provides 
a framework to calculate the quantum amplitude for every number of spatial dimensions
consistent with string theory. The probability distribution of various dimensions for the universe 
is of little significance, however, because we have already measured we live in four dimensions. 
Our observation only gives us a single number, so we cannot tell from this whether the universe 
was likely to be four dimensional, or whether it was just a lucky chance. Hence as long as the no boundary 
amplitude for three large spatial dimensions is not exactly zero, the observation that $D=3$ does not help to 
prove or disprove the theory. Instead of asking for the probabilities of various dimensions for the 
universe, therefore, we might as well use our observation as a final boundary condition and consider only 
amplitudes for surfaces $S$ with three large dimensions. The number of dimensions is thus best used as a 
constraint to restrict the class of histories that contribute to the path integral for a universe like ours. This
restriction allows one to identify definite predictions for future observations.

The situation with the low energy effective theory of particle interactions may well be similar.
In string theory this is the effective field theory for the modular parameters that describe the 
internal space. It is well known that string theory has solutions with many different compact
manifolds. The corresponding effective field theories are determined by the topology and the geometry
of the internal space, as well as the set of fluxes that wind the 3-cycles. Furthermore, for each 
effective field theory the potential for the moduli typically has a large number of local minima. Each 
local minimum of the potential is presumably a valid vacuum of the theory. These form a landscape 
\cite{Susskind03} of possible stable or metastable states for the universe at the present time, each with 
a different theory of low energy particle physics. 

In the bottom-up picture it is thought that the universe begins with a grand unified symmetry, such as 
$E_8 \times E_8$. As the universe expands and cools the symmetry breaks to the Standard Model, perhaps 
through intermediate stages. The idea is that string theory predicts the pattern of breaking, and the 
masses, couplings and mixing angles of the Standard Model. However, as with the dimension of space, there 
seems to be no particular reason why the universe should evolve precisely to the internal space that gives 
the Standard Model\footnote{An extension of the bottom-up approach invokes the notion of eternal inflation 
to accomodate the possibility that the position in the moduli space falls into different minima in different
places in space, leading to a mosaic structure for the universe. The problem with this approach is that one 
cannot predict what a typical local observer within such a universe would see. We discuss this in more 
detail in Section 7.}. It is therefore more useful to compute no boundary amplitudes for a spacelike 
surface $S$ with a given internal space. This is the top down approach, where one sums only over the 
subclass of histories which end up on $S$ with the internal space for the Standard Model. 

We now turn to the predictions $\alpha$ we can expect to derive from amplitudes like (\ref{condprob}). 
We have seen that the relative amplitudes for radically different geometries are often irrelevant. By 
contrast, the probabilities for neighbouring geometries are important. The most powerful predictions are 
obtained from the relative amplitudes of nearby geometries, conditioned on various discrete 
features of the universe. This is because these amplitudes are not determined by the selection effects of 
the final boundary conditions. Rather, they depend on the quantum state $\vert \Psi \rangle$ itself. 

Neighbouring geometries correspond to small quantum fluctuations of continuous quantities, like the temperature of 
the cosmic microwave background (CMB) radiation or the expectation values of the string theory moduli in a given 
vacuum. In inflationary universes these fluctuations are amplified and stretched, generating a pattern of spatial 
variations on cosmological scales in those directions of moduli space that are relatively flat\footnote{Spatial 
variations of coupling constants from scalar moduli field fluctuations generate large scale isocurvature fluctuations 
in the matter and radiation components \cite{Kofman03}.}.
The spectra depend on the quantum state of the universe.  Correlators of fluctuations in the no boundary state can 
be calculated by perturbatively evaluating the path integral around instanton saddle points \cite{Gratton99}. In 
general if ${\cal P}(x_1)$ and ${\cal Q}(x_2)$ are two observables at $x_1$ and $x_2$ on a final surface $S$, then 
their correlator is formally given by the following integral over a complete set of observables ${\cal O}(x)$ on 
$S$ \cite{Gratton99},
\be\label{corr}
\langle  {\cal P}(x_1){\cal Q}(x_2) \rangle \sim
\sum_B \int [{\cal D O}(S)] \Psi_B [{\cal O}]^* \Psi_B [{\cal O}]{\cal P}(x_1){\cal Q}(x_2). 
\ee
Here the sum is taken over backgrounds $B$ that satisfy the prescribed conditions on $S$. The amplitude 
$\Psi_{B}$ for fluctuations about a particular background geometry $(\bar g, \bar \phi)$ is given by
\be\label{nbpfluct}
\Psi_B [g^3,\phi ] \sim e^{-S_0(\bar g,\bar \phi)}
\int [{\cal D} \delta g ][{\cal D} \delta \phi] \ e^{ -S_{2}[\delta g,\delta \phi]}
\ee
where the metric $g=\bar g + \delta g$ and the fields $\phi = \bar \phi +\delta \phi$. 
The $C_{l}$'s of the CMB temperature anisotropies are classic examples of observables that can be calculated from 
correlators like this. Whilst the full correlator (\ref{corr}) generally involves a sum over several saddle points, 
for most practical purposes only the lowest action instanton matters. 

In no boundary backgrounds like the HM geometry, where a real Euclidean instanton is matched onto a real Lorentzian 
metric, one can find the correlators by first calculating the 2-point functions in the 
Euclidean region. The Euclidean correlators are then analytically continued into the Lorentzian region, where they 
describe the quantum mechanical vacuum fluctuations of the various fields in the state determined by no boundary 
initial conditions. The path integral unambiguously specifies boundary conditions on the Euclidean fluctuation modes. 
This essentially determines a reflection amplitude $R(k)$, where $k$ is the wavenumber, which depends on the instanton 
geometry. The spectra in the Lorentzian, and in particular the primordial gravitational wave spectrum 
\cite{Gratton00}, depend on the instanton background through $R(k)$.

The relative amplitudes of neighbouring geometries can thus be used to predict, from first principles, the 
precise shape of the primordial fluctuation spectra that we observe. This provides a test of the no boundary proposal 
and, more generally, an observational discriminant between different proposals for the state of the universe, because 
the spectra contain a signature of the initial conditions. 

Before we illustrate the top down approach in a simple model in Section 5, we briefly comment on the role of anthropic 
selection effects in top down cosmology.

%%%%%%%%%%%%%%%%%%%%%%%%%%%%%%
\section{Anthropic Reasoning}
%%%%%%%%%%%%%%%%%%%%%%%%%%%%%%

In general anthropic reasoning \cite{Barrow86}
aims to explain certain features of our universe from our existence in it. 
One possible motivation for this line of reasoning is that the observed values and correlations of certain 
parameters in particle physics and cosmology appear necessary to ensure life emerges in our universe. If 
this is indeed the case it seems reasonable to suppose that certain environmental selection effects 
need to be taken in account in the calculation of probabilities for observations.

It has been pointed out many times, however, that anthropic reasoning is meaningless if it is not 
implemented in a theoretical framework that determines which parameters can vary and how they vary. Top down 
cosmology, by combining the string landscape with the no boundary proposal, provides such a 
framework\footnote{Several alternatives to this framework have been proposed, and we comment on some of these 
in Section 7.}. The anthropic principle is implemented in the top down approach by specifying a set of 
conditions $\beta$ in (\ref{condprob}) that select the subclass of histories where life is likely to emerge.
More specifically, anthropic reasoning in the context of top down cosmology amounts to the evaluation
of conditional probabilities like
\be\label{anthr}
p(\alpha \vert O, H, \Psi),
\ee
where $O$ represents a set of conditions that are required for the appearance of complex 
life. The utility and predictivity of anthropic reasoning depends on how sensitive the probabilities 
(\ref{anthr}) are to the inclusion of $O$. Anthropic reasoning is useful and predictive only if 
(\ref{anthr}) is sharply peaked around the observed value of $\alpha$, and if the {\it a priori} theoretical 
probability $p(\alpha \vert H,\Psi)$ itself is broadly distributed \cite{Hartle05}. 

Anthropic reasoning, therefore, can be naturally incorporated in the top down approach. In particular it may 
provide a qualitative understanding for the origin of certain conditions $\beta$ that one finds are 
useful in top 
down cosmology. Consider the number of dimensions of space, for example. We have argued that this is best 
used as a final constraint, but the top down approach itself does not explain why this particular property of the 
universe 
cannot be predicted from first principles. In particular, the top down argument does not depend on whether 
four dimensions is the only arena for life. Rather, it is that the probability distribution over dimensions 
is irrelevant, because we cannot use our observation that $D=3$ to falsify the theory. But it may turn 
out that anthropically weighted probabilities (\ref{anthr}) are always sharply peaked around $D=3$.
In this case one 
can essentially interpret the number of dimensions as an anthropic requirement, and it would be an example where
anthropic reasoning is useful 
to understand why one needs to condition on the number of dimensions in top down cosmology.

We emphasize, however, that the top down approach developed here goes well beyond conventional anthropic reasoning. 
Firstly, the top down approach gives {\em a priori} probabilities that are more sharply peaked, because it adopts 
a concrete prescription for the quantum state of the universe - as opposed to the usual assumption that predictions 
are independent of $\Psi$. Hence the framework we propose is more predictive than conventional anthropic 
reasoning\footnote{Anthropic selection effects have been used to constrain the value of the
cosmological constant \cite{Weinberg87}, and the dark matter density \cite{Hellerman05}. In these studies
it is assumed, however, that the a priori probability distributions are independent of the state of the 
universe. This reduces the predictivity of the calculations, and could in fact be misleading.}.

Top down cosmology is also more general than anthropic reasoning, because there is a wider range of 
selection effects that can be quantitatively taken in account. In particular the conditions $\beta$ that 
are supplied in (\ref{condprob}) need not depend on whether they are necessary for life to emerge. The set of 
conditions generally includes environmental selection effects similar to anthropic requirements, but it also 
includes chance outcomes of quantum accidents in the early universe that became frozen. The latter need not be
relevant to the emergence of life. Furthermore, they cannot be taken in account by simply adding an 
{\em a posteriori} selection factor proportional to the number density of some reference object, because 
they change the entire history of the universe! 

We illustrate this in the next section, where we derive several predictions of top down cosmology
in a simple toy model.

%%%%%%%%%%%%%%%%%%%%%%%%%%%%%%
\section{Models of Inflation}
%%%%%%%%%%%%%%%%%%%%%%%%%%%%%%

How can one get a nonzero amplitude for the present state of the universe if, as we claim, the 
metrics in the sum over histories have no boundary apart from the surface $S$ at the present time?
We do not have a definitive answer, but one possibility would be if the four dimensional part of the 
saddle point metric was an inflating universe at early times. Hartle and Hawking \cite{Hartle83}
have shown that such metrics 
can be rounded off in the past, without a singular beginning and with curvature bounded well away the Planck 
value.  They give a nonzero value of the no boundary amplitude for almost any universe that arises from an 
early period of inflation. Thus to illustrate the top down approach described above, we consider a simple model with 
a few positive extrema of the effective potential.

We assume the instability of the inflationary phase can be described as the evolution of a 
scalar order parameter $\phi$ moving in a double well potential $V(\phi)$, shown in Figure 1. 
We take the potential to have a broad flat-topped maximum $V_0$ at $\phi=0$ and a minimum at 
$\phi_1$. The value at the bottom is the present small cosmological constant $\Lambda$. A concrete 
example would be gravity coupled to a large number of light matter fields \cite{Starobinsky80}. 
The trace anomaly generates a potential which has unstable de Sitter space as a self-consistent 
solution\footnote{See \cite{Hawking01} for an earlier discussion of trace anomaly inflation with 
no boundary initial conditions.}.

%%%%%%%%%%%%%%%%%%%%%%%%%%%%%%%
\begin{figure}[htb]
\begin{picture}(0,0)
\put(120,145){$V$}
\put(120,116){$V_0$}
\put(120,20){$\Lambda$}
\put(209,15){$\phi_{1}$}
\put(235,10){$\phi$}
\end{picture}
\centering{\psfig{file=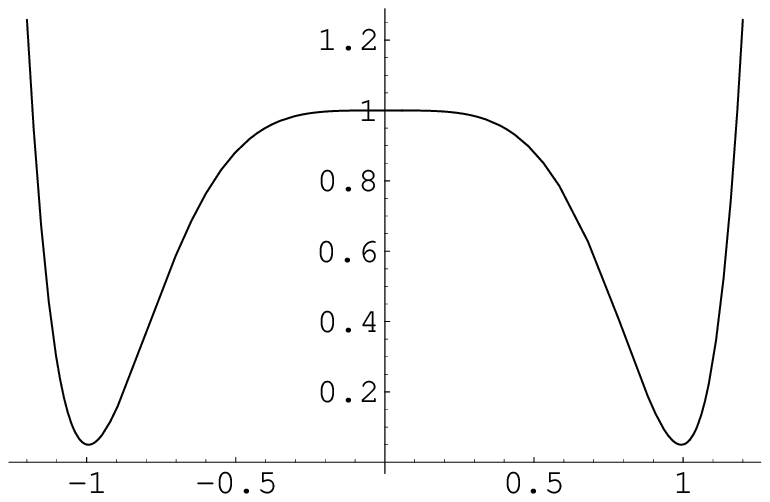,width=3.2in}}
\caption{}
\label{1}
\end{figure}
%%%%%%%%%%%%%%%%%%%%%%%%%%%%%%%%%%%

We are interested in calculating the no boundary amplitude of an expanding non-empty region 
of spacetime similar to the one we observe today. In the semi-classical approximation, this will come from one or 
more saddle points in the action. These correspond to solutions of the Einstein equations. One solution is de Sitter 
space with the field $\phi$ sitting at the minimum of the potential $V(\phi)$. This will have a very large amplitude, 
but will be complete empty and therefore does not contribute to the top down amplitude for a universe like ours. To 
obtain an
expanding universe with $\Omega_{m} \sim {\cal O}(1)$ and with small perturbations that lead to galaxies, it seems 
necessary to have a period of inflation
\footnote{One might think it would be more likely for a universe like ours to arise from a fluctuation of the big 
de Sitter space directly into a hot big bang, rather than from a homogeneous fluctuation up the 
potential hill that leads to a period of inflation. The amplitude of a hot big bang fluctuation is much smaller, 
however, than the amplitude of the inflationary saddle points we discuss below
(see also \cite{Albrecht04}). The latter do not directly connect to the large de Sitter space, but they could be 
connected with very little cost in action by a thin bridge \cite{Hawking84}.}.

We therefore  consider the no boundary amplitude\footnote{We work in the $K$ representation of the wave function 
(see e.g. \cite{Hawking84}), where one replaces $g^3_{ij}$ on the three-surface $S$ by $\tilde g^3_{ij}$, the 
three-metric up to a conformal factor, and $K$, the trace of the second fundamental form.
The action $S_{E}^{k}$ differs from (\ref{Eact}) in that the surface term has a coefficient 1/3.} 
 $\Phi [\tilde g^3,K,\phi ]$ for a closed inflating universe bounded by
a three-surface $S$ with a large approximately constant Hubble parameter $H =\dot a/a$ (and corresponding 
trace $K=-3\dot a/a=-3H$), and a nearly constant field $\phi$ near the top of $V$. The value of $\phi$ on $S$ is 
chosen sufficiently far away from the minimum of $V$ to ensure there are at least enough efoldings of inflation
for the universe at the present time to be approximately flat. 

We first calculate the wave function for imaginary $K$, or real Euclidean $K_{e}=iK$, and then analytically 
continue the result to real Lorentzian $K$. There are two distinct saddle point contributions to the amplitude 
for an inflating universe in this model \cite{Hawking02}. In the first case, the universe is created by the HM 
instanton with constant $\phi=0$. Then quantum fluctuations disturb the 
field, causing it to classically roll down the potential to its prescribed value on $S$. Histories of this kind 
thus have a long period of inflation, and lead to a perfectly flat universe today. The action of the HM geometry 
is given by 
\ba \label{act1}
S_{HM}^{k} (K) = - \frac{12\pi^2}{V_0}
\left(1-\frac{ K_{e}}{(V_0^2+ K^2_e)^{1/2}}\right)
\ea
where $K_{e} =3b_{,\tau}/b$.

There is, however, a second saddle point contribution which comes from a deformed four sphere, with line element 
\be \label{def}
ds^2 = d\tau^2 +b^2 (\tau) d\Omega^2_3,
\ee
where $\phi (\tau)$ varies across the instanton. The Euclidean field equations for $O(4)$-invariant instantons are
\be \label{efield1}
\phi'' =  -K_e\phi' +V_{,\phi},
\qquad \qquad 
K_{e}' +K^2_{e}=  -(\phi^2_{,\tau} +V )
\ee
where $\phi'=\phi_{,\tau}$. 
These equations admit a solution, which is part of a Hawking--Turok instanton\footnote{There is no CdL instanton that 
straddles 
the maximum in our model, because we have assumed the potential has a broad flat-topped maximum, 
$\vert V''(0) \vert /H^2 \leq 1$.} \cite{Hawking98}, where $\phi$ 
slowly rolls up the potential from some value $\phi_0$ at the (regular) South Pole to its prescribed value on the 
three-surface $S$. Hence this solution represents a class of histories where the scalar starts as far down 
the potential as the condition that the present universe be approximately flat allows it to. This naturally leads to 
fewer efoldings of inflation, 
and hence a universe that is only approximately flat today. The Euclidean action $S_{HT}^{k}(K)$ of the 
deformed four sphere was given in \cite{Hawking02} (eq.4.8), in the approximation that $\phi$ is reasonably 
small everywhere.

A comparison of the action of both saddle points shows that the deformed four sphere dominates the path integral 
for amplitudes with real Euclidean $K_{e}$ on $S$. This would seem to suggest that the universe is least likely 
to start with $\phi$ at the top of the hill. However, we are interested in the amplitude for an expanding 
Lorentzian universe, with real Lorentzian $K$ on $S$. If one analytically continues the action into the complex 
$K_{e}$-plane, one finds the action of the deformed four sphere rapidly increases along the imaginary $K_{e}$-
axis whereas the real part of $S_{HM}^{K}$ remains constant, and the dominant contribution to amplitudes for larger  
$K$ on $S$ actually comes from the HM geometry. The reason for this is that a constant scalar field saves 
more in gradient energy, than it pays in potential energy for being at the top of the hill. 
Hence a Lorentzian, expanding universe with large Hubble parameter $H$ is most likely to emerge in an inflationary 
state, with $\phi$ constant at the maximum of the potential. 

Top down cosmology thus predicts that in models like trace anomaly inflation, expanding universes with small 
perturbations that lead to galaxies, start with a long period of inflation, and are perfectly flat today.
Furthermore, as discussed earlier, the precise shape of the primordial fluctuation spectra can be computed from the 
Euclidean path integral, by perturbatively evaluating around the HM saddle point.

%%%%%%%%%%%%%%%%%%%%%%%%%%%%
\section{Prediction in a Potential Landscape}
%%%%%%%%%%%%%%%%%%%%%%%%%%%%

The predictions we obtained in the previous section extend in a rather obvious way to models where one has
a potential landscape. A generic potential landscape admits a large class of alternative inflationary histories 
with no boundary initial conditions. 
There will be HM geometries at all positive saddle points of the potential. For saddle points with more than one 
descent direction, there will generally be a lower saddle point with only one descent direction, and with 
lower action.  
If this descent direction is sharply curved $\vert V''(0) \vert /H^2 > 1$, one would not expect a significant 
top down amplitude to come from the saddle point. Thus only broad saddle points with a single descent direction 
will give 
rise to amplitudes for universes like our own. The requirement\footnote{Extra constraints from particle physics, 
when combined with the cosmological constraints discussed here, will probably further raise the value of $V_0$.}
that the primordial fluctuations be sufficiently 
large to form galaxies, however, sets a lower bound on the value of $V_0$. 

Only a few of the saddle points will satisfy the demanding condition that they be broad, because it requires 
that 
the scalar field varies by order the Planck value across them. Because the dominant saddle points are in the 
semi-classical regime, the solutions will evolve from the saddle points to the neighbouring minima of $V$. 
Thus top down cosmology predicts that only a few of the possible vacua in the landscape will have significant 
amplitudes.

%%%%%%%%%%%%%%%%%%%%%%%%%%%%%%
\section{Alternative Proposals}
%%%%%%%%%%%%%%%%%%%%%%%%%%%%%%

To conclude, we briefly comment on a number of different approaches to the problem of initial conditions in 
cosmology, and we clarify in what respect they differ from the top down approach we have put forward\footnote{We 
believe the 
framework described here addresses the concerns raised in \cite{Aguirre05} regarding a top down 
approach cosmology.}.

We have already discussed the pre-big bang cosmologies \cite{Veneziano91}, where one specifies initial 
conditions in the infinite past and follows forward in time a single semi-classical history of the universe. 
Pre-big bang cosmology is thus based on a bottom-up approach to cosmology. It requires one to postulate 
a fine-tuned initial state, in order to have a smooth deterministic transition through the 
big crunch singularity.

We have also discussed the anthropic principle \cite{Barrow86}. This can be implemented in top down 
cosmology, through the specification of final boundary conditions that select histories where life emerges.
Anthropic reasoning within the top down approach is reasonably well-defined, and useful to the extent that it 
provides a qualitative understanding for the origin of certain late time conditions that one finds 
are needed in top down cosmology.

%%%%%%%%%%%%%%%%%%%%%%%%%%%%%%
\subsection{Eternal Inflation}
%%%%%%%%%%%%%%%%%%%%%%%%%%%%%%

A different approach to string cosmology has been to invoke the phenomenon of eternal inflation 
\cite{Vilenkin83} to populate the 
landscape. There are two different mechanisms to drive eternal inflation, which operate in 
different moduli space regions of the landscape. In regions where the moduli potential monotonically 
increases away from its minimum, it is argued that inflation can be sustained forever by quantum 
fluctuations up the potential hill. Other regions of the landscape are said to be 
populated by the nucleation of bubbles in metastable de Sitter regions. The interior of these bubbles may 
or may not exit inflation, depending on the shape of the potential across the barrier. 

Both mechanisms of eternal inflation lead to a mosaic structure for the universe, where causally disconnected
thermalized regions with different values for various effective coupling constants are separated from each other 
by a variety of inflating patches. It has proven difficult, however, to calculate the probability distributions 
for the values of the constants that a local observer in an eternally inflating universe would 
measure\footnote{See however \cite{Garriga05} for recent progress on this problem.}. This is
because there are typically an infinite number of thermalized regions.

One could also consider the no boundary amplitude for universes with a mosaic structure.
However, these amplitudes would be much lower than the amplitudes for final states that are homogeneous 
and lie entirely within a single minimum, because the gradient energy in a mosaic universe contributes positively
to the Euclidean action. 
Histories in which the universe eternally inflates, therefore, hardly contribute to the no boundary 
amplitudes we measure. Thus the global structure of the universe that eternal inflation predicts, 
differs from the global structure predicted by top down cosmology. Essentially this is because eternal 
inflation is again based on the classical idea of a unique 
history of the universe, whereas the top down approach is based on the quantum sum over histories. The key 
difference between both cosmologies is that in the proposal based on eternal inflation there is thought to be 
only one universe with a fractal structure at late times, whereas in top down cosmology one envisions a set of 
alternative universes, which are more likely to be homogeneous, but with different values for various 
effective coupling constants.

It nevertheless remains a challenge to identify predictions that would provide a clear observational 
discriminant between both proposals\footnote{It has been argued \cite{Freivogel05} that eternal inflation in 
the string landscape predicts we live in an open universe. It seems this is not a prediction of no boundary initial 
conditions on the string landscape; the HM geometries we discussed occur generically and thus 
provide a counterexample.}.
We emphasize, however, that even a precise calculation of conditional probabilities in no boundary 
cosmology, which takes in account the backreaction of quantum fluctuations, will make no reference to the 
exterior of our past light cone. Indeed, the top down framework we have put forward indicates that the 
mosaic structure of an eternally inflating universe is a redundant theoretical construction, which should be 
excised by Ockham's razor\footnote{Or on the basis of holography \cite{Banks05}?}. It appears unlikely, 
therefore, that something like a `volume-weighted' probability distribution - which underlies the idea of 
eternal inflation - can arise from calculations in top down cosmology. The implementation of
selection effects in both approaches is fundamentally different, and this should ultimately translate into 
distinct predictions for observations.

%%%%%%%%%%%%%%%%%%%%%%%%%%%%%%
\section{Concluding Remarks}
%%%%%%%%%%%%%%%%%%%%%%%%%%%%%%

In conclusion, the bottom up approach to cosmology would be appropriate, if one knew that the universe was set 
going in a particular way in either the finite or infinite past. However, in the absence of such 
knowledge one is required to work from the top down. 

In a top down approach one computes amplitudes for alternative histories of the universe with final boundary 
conditions only. The boundary conditions act as late time constraints on the alternatives and select the subclass 
of histories that contribute to the amplitude of interest. This enables one to test the proposal, by searching
among the conditional probabilities for predictions of future observations with probabilities near one.
In top down cosmology the histories of the universe thus depend on the precise question asked, i.e. on the set of 
constraints that one imposes. There are histories in which the universe eternally inflates, or is eleven dimensional, 
but we have seen they hardly contribute to the amplitudes we measure. 

A central idea that underlies the top down approach is the interplay between the fundamental laws of nature and 
the operation of chance in a quantum universe. In top down cosmology, the structure and complexity of alternative 
universes in the landscape is predictable from first principles to some extent, but also 
determined by the outcome of quantum accidents over the course of their histories.

We have illustrated our framework in a simple model of gravity coupled to a scalar with a double well potential, and
a small fundamental cosmological constant $\Lambda$. Imposing constraints that select the subclass of histories 
that are three dimensional and approximately flat at late times, with sufficiently large primordial perturbations 
for structure formation to occur, we made several predictions in this model.

In particular we have shown that universes within this class are likely to emerge in an inflationary state.
Furthermore, we were able to identify the dominant inflationary path as the history where the scalar starts all the
way at the maximum of its potential, leading to a long period of inflation and a perfectly flat universe today.
Moreover, one can calculate the relative amplitudes of neighbouring geometries by perturbatively evaluating
the path integral around the dominant saddle point. Neighbouring geometries correspond to small 
quantum fluctuations of various continuous quantities, like the temperature of the CMB radiation or the expectation 
values of moduli fields. In inflationary universes these fluctuations are amplified and stretched, 
generating a pattern of spatial variations on cosmological scales in those directions of moduli space that 
are relatively flat. The shape of these primordial spectra depends on the (no) boundary conditions on the
dominant geometry and provides a strong test of the no boundary proposal.

When one extends these considerations to a potential that depends on a multi-dimensional moduli space, one finds 
that only a few of the minima of the potential will be populated, i.e. will have significant amplitudes.

The top down approach we have described leads to a profoundly different view of cosmology, and the relation between 
cause and effect. Top down cosmology is a framework in which one essentially traces the histories backwards, from a 
spacelike surface at the present time. The no boundary 
histories of the universe thus depend on what is being observed, contrary to the usual idea that the universe has 
a unique, observer independent history. In some sense no boundary initial conditions represent a sum over all possible 
initial states. This is in sharp contrast with the bottom-up approach, where one assumes 
there is a single history with a well defined starting point and evolution. 
Our comparison with eternal inflation provides a clear illustration of this.
In a cosmology based on eternal inflation there is only one universe with a fractal structure at late times, 
whereas in top down cosmology one envisions a set of alternative universes, which are more likely to be homogeneous, 
but with different values for various effective coupling constants.

%\bigskip

\centerline{{\bf Acknowledgments}}

We thank Jim Hartle for valuable and stimulating discussions over many years.

\end{document}